\documentclass[prb,twocolumn,showpacs,amsmath,amssymb]{revtex4-1}
\usepackage{graphicx,amsmath,bm, braket, txfonts, color}
\pdfoutput=1

\begin{document}

\title{Giant magnetoelectric effect in  magnetic tunnel junctions
coupled to an electromagnetic environment}
\author{Mircea Trif}
\author{Pascal Simon}
\affiliation{Laboratoire de Physique des Solides, CNRS, Universit\'{e} Paris-Sud, 91405 Orsay, France}
 
\date{\today}

\begin{abstract}
We study the magnetization dynamics in ferromagnet$\mid$insulator$\mid$ferromagnet and ferromagnet$\mid$insulator$\mid$normal metal ultra-small tunnel junctions, and the associated voltage drop in the presence of an electromagnetic environment assisting the tunneling processes. We show that the environment strongly affects the resulting voltage, which becomes a highly non-linear function of the precession cone angle $\theta$.  We find  that voltages comparable to the driving frequency $\omega$ can be reached  even for small precession cone angles $\theta$, in stark contrast to the case where the  environment is absent. Such an effect could be useful for the detection local magnetization precessions in textured ferromagnets or, conversely,  for probing  the environment via the magnetization dynamics.     
\end{abstract}


\maketitle

\section{Introduction}
\label{sec1}

Magnetic tunnel junctions (MTJs) are usual tunnel junctions where  two metallic
ferromagnets are separated by a small insulating barrier \cite{IkedaIEEE07}. In MTJs the tunneling current depends on
the relative orientation of the magnetizations of the two ferromagnetic layers which can be controlled
by external magnetic fields \cite{ModeraPRL95}, by electrical currents \cite{RalphJMMM08,MaekawaBook}, or even by voltages \cite{MaruyamaNatNano09,ZhuPRL2012}. This phenomenon is called tunneling magnetoresistance (TMR) and it is a consequence of the spin-dependent electron tunneling. There is a plethora of applications utilizing this effect such as,  for example, 
magnetic random access memories and magnetic sensors \cite{TehraniIEEE99}.

Miniaturization lies at the heart of both electronics and  spintronics (electronics with spin). The drawback, however, is that the  electronic or spintronic elements become more vulnerable to parasitic effects caused by the surrounding noise sources. For example, by reducing the size of a  tunnel junction, the surrounding electromagnetic environment becomes manifest, strongly affecting the electronic transport \cite{IngoldNazarov1990}. One such effect is the dynamical Coulomb blockade (DCB), which  means the  activation of a Coulomb gap at finite voltage due to  inelastic tunneling processes assisted  by the environment the tunnel junction interacts with \cite{DevoretPRL90,GirvinPRL90}. Along with this effect there are many other physical  consequences, such as the zero bias anomaly, single-electron tunneling (SET) oscillations, etc,  and for more details we refer the reader to Refs.~\onlinecite{IngoldNazarov1990, BlanterNazarov2009}.  
 
Similarly, by reducing the size of a MTJ one expects that
 Coulomb interaction effects  become important in the presence of an environment
(there is always  the electronics surrounding the MTJ that can act as an environment, or even the intrinsic electron-electron interactions within the material). Even though typical environments are not spin-dependent, they can still affect the spin-dependent transport, and thus quantities such as the magnetoconductance and current induced spin-transfer torques. One important 
 application of MTJs is that they can be efficiently used as magnetization detectors: if one of the
two magnet is driven at its ferromagnetic resonance, this results in a dc current flowing
through the junction which, in the open circuit, translates into a voltage buildup \cite{XiaoPRB08,MoriyamaPRL08,TserkovnyakPRB08}. For example, by scanning the surface of a textured magnet with the
help of a spin-polarized STM tip, one can detect the local magnetization dynamics instead of the global one 
(e.g. rotation of skyrmions, magnetic vortices, magnetic spirals, etc) \cite{NagaosaNatNano13,SchulzNatPhys12,SampaioNatNano13,OkamuraNatCom13}. The disadvantage of such a
detection scheme is that although the signal is proportional to the precession frequency $\omega$, it is also
proportional to the square of the precession angle (or cone angle) $\theta$, which is typically  very small, of
about $1$-$5$ degrees \cite{TserkovnyakRevModPhys06}. Thus, the conversion of magnetization dynamics into a voltage is rather
inefficient, although some experimental evidences have been put forward \cite{MoriyamaPRL08,YamanePRL11}. 
\begin{figure}[t]
\begin{center}
\includegraphics[width=0.9\linewidth]{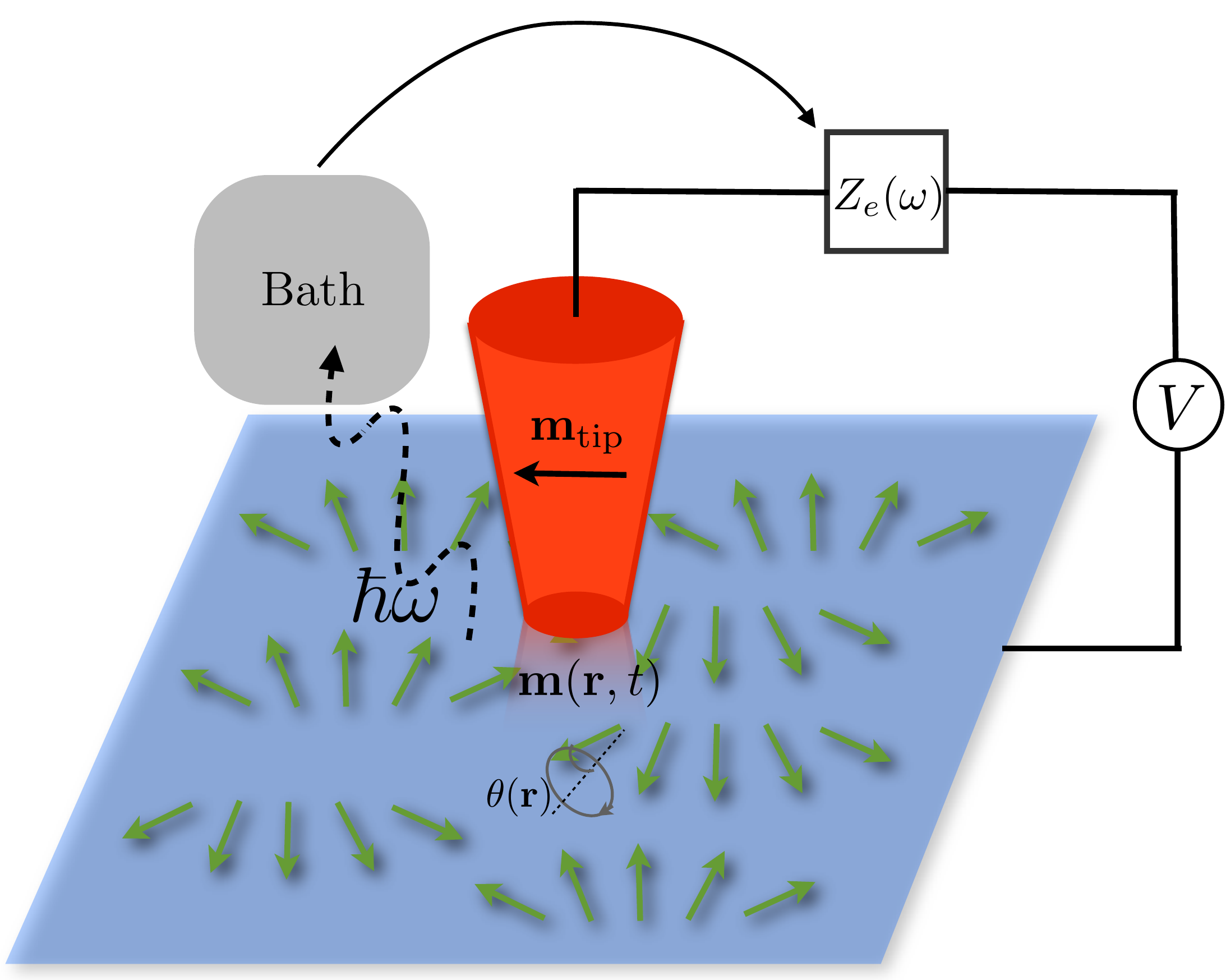}
\caption{Sketch of the system. The textured ferromagnet (in blue) with a time-dependent magnetization field $\bm{m}(\bm{r},t)$ (the green arrows) precessing at a cone angles $\theta(\bm{r})$ is coupled to a  static ferromagnet on the right side with magnetization $\bm{m}_{\rm tip}\equiv\bm{m}_{R}$ by a small tunnel junction of capacitance $C$. Here, the tunneling process  can be inelastic,  i.e. assisted by excitations $\hbar\omega$ in an external bath (depicted in gray) that  affects the resulting electrical current $I$ in the circuit. The bath (or environment) is modeled by a frequency-dependent impedance $Z_e(\omega)$ in series with the base circuit. In the open circuit, there is a voltage drop build-up $V$ that can be accessed with a voltmeter.}
\label{STM_sketch}
\end{center}
\end{figure}
There are several proposals on how to increase such a detection scheme  by resorting to various nonlinearities, either by using quantum dots to explore the (static) Coulomb blockade \cite{BenderPRB10}, or even by using superconductors in order to  take advantage of their singular densities of states \cite{trifPRL13}. In this paper we take a step forward and analyze the original  \cite{MoriyamaPRL08}, but in the presence of an environment assisting the tunneling processes. 

The paper is organized as follows. In Sec.~\ref{sec2} we introduce the setup and the system Hamiltonian in the presence of the magnetization dynamics. In Sec.~\ref{sec3} we calculate the electronic tunneling rates and the charge currents flowing through the MTJ due to the precessing magnetization. In Sec.~\ref{sec4} we investigate the magnetization induced voltage in the open circuit for two types of environments, namely single-mode and ohmic, respectively. We show that the coupling to the environment leads, in both cases, to a singular behavior in the electronic transport that dramatically enhances the induced voltage. Here, we also discuss the effect of thermal fluctuations on the induced signal, and demonstrate that the singular features persist at low temperatures. Finally, in Sec.~\ref{sec5} we end up with some conclusions and outlook on the possible utilization of the effect. 
%We show that this strongly affects the charge currents, and  that it can  lead to a dramatic enhancement of  the voltage buildup in the open circuit, reaching values of the order of  the driving frequency $\omega$ for small precession cone angles $\theta$.  

\section{System Hamiltonian}
\label{sec2}

In Fig.~\ref{STM_sketch} we show a sketch of the system under consideration which consist of a spin-polarized STM tip scanning a (dynamical) textured ferromagnet in the presence of inelastic processes. In the following, however,  we assume the tip is fixed at some position $\bm{r}_{\rm tip}$, so that $\bm{m}(\bm{r}_{\rm tip},t)\equiv \bm{m}_{L}(t)$, $\theta(\bm{r}_{\rm tip})\equiv\theta$, and $\omega(\bm{r}_{\rm tip})\equiv\omega$, while $\bm{m}_{\rm tip}\equiv \bm{m}_R$. The  Hamiltonian describing the electrons in the the leads (sample and tip) reads:
\begin{equation}
H_{\rm F}=\frac{p^2}{2m}+V(\bm{r})+\sum_{i=L,R}\frac{\Delta_i}{2}\bm{m}_i(\bm{r},t)\cdot\bm{\sigma}\,,
\end{equation} 
with $i=L, R$ (left, right) leads,  $\bm{m}_i(\bm{r},t)$ is the unit vector pointing along the instantaneous magnetization direction in lead $i$,  $V(\bm{r})$ includes any electric-field like potentials, including disorder, external electric fields, crystal, etc, and $\Delta_i$ is the exchange splitting in the lead $i=L$, $R$. Note that for a normal right lead (normal metal tip) we have $\Delta_R=0$. The magnetization direction in the left lead read $\bm{m}_L(t)=(\cos{\omega t}\sin{\theta},\sin{\omega t}\sin{\theta},\cos{\theta})$, namely it precesses at frequency $\omega$ and cone angle $\theta$ around the $\hat{z}$-axis. The right ferromagnet instead has its magnetization direction pointing along $z$, i.e. $\bm{m}_R=\hat{\bm{z}}$. 

The leads Hamiltonian needs to be supplemented  by the tunneling term \cite{DevoretPRL90}: 
\begin{align}
H_{\rm T}&=\sum_{kq;\sigma\sigma'}T^{kq}_{\sigma\sigma'}c^{\dagger}_{k\sigma}c_{q\sigma'}\Lambda_e+{\rm h. c.}\,,
\end{align}
where $k\sigma$ $(q\sigma')$ are  the momentum and spin in the left (right) lead respectively, $T^{kq}_{\sigma\sigma'}$ is the spin-dependent tunneling matrix element, and $\Lambda_e$ is an operator that changes the number of electrons $Q$ on the plates of the tunnel junction by one, i.e. its effect can be formally written as  $\Lambda_eQ\Lambda_e^\dagger=Q-e$. We assume that the spin-orbit interaction is absent, so that the only spin dependence of the tunneling matrix elements arise because  of the finite magnetization in the leads. Assuming that $T^{kq}_{\sigma\sigma'}$ does not depend on momentum (as all transport happens around  the Fermi level), we are left with the following  matrix element:
\begin{equation}
T^{kq}_{\sigma\sigma'}\equiv t\langle\sigma|\sigma'\rangle=t\left(\delta_{\sigma\sigma'}\cos{\frac{\theta}{2}}+\delta_{\sigma\bar{\sigma}'}\sin{\frac{\theta}{2}}\right)\,,
\end{equation}
where, however, the two spin states are represented in a basis corresponding to the left and right leads magnetizations, respectively. We note that the ``displacement" operator is given by  $\Lambda_e=e^{i\phi}$, where $\phi$ is a phase operator  which is conjugate with the charge on the capacitor, i.e. $[Q,\phi]=ie$ \cite{IngoldNazarov1990}. The dynamics of the phase $\phi$ is dictated by the environment the junction is embedded in, and  different environments will be discussed  later on. For starters, we only assume that the angle $\phi$ is a dynamical variable.  

The electronic transport in the presence of the magnetization dynamics can be easily tackled  by resorting to the rotating wave description, meaning the time-dependent setup can be transformed into a static one. Consequently, we can use usual static tunneling description, such as the Fermi's golden rule, in order to calculate the charge currents. We will forget altogether about any spin-related processes such as spin pumping and spin accumulations since tunneling times are assumed to be much longer than the spin relaxation times in typical ferromagnetic leads.  We then perform the following unitary transformation:
 \begin{equation}
 H_{\rm RW}(t)\rightarrow U^\dagger(t) H_{\rm tot}(t)U(t)-iU^\dagger(t)\partial_tU(t)=H_{\rm tot}(0)-(\omega/2)\sigma_z\,,
 \end{equation}
where $H_{\rm tot}=H_{F}+H_T$, and  $U(t)=\exp{(-i\omega t\sigma_z)}$, so that we formally eliminated the time-dependence on the left lead. This results in a shift of the energy levels on the right lead by $\pm\omega/2$ for spin up (down), while in the left  we obtain the following Hamiltonian \cite{TserkovnyakPRB08}:
\begin{equation}
H_L(t)\rightarrow H_L(0)-(\omega/2)\cos{\theta}\sigma_{\parallel}\,.
\end{equation}
Note that $U(t)$ commutes with any charge degree of freedom, and thus this transformation leaves the environment unaffected. In Fig.~\ref{Levels_sketch} we show the resulting levels structure in the rotating frame and the fictitious spin splittings in the the two parts of the setup.  Let us provide now the physical picture behind this transformation: in the rotating frame there are static spin biases emerging (or spin-motive forces), which can drive  charge currents. For that to occur, there should be spin decaying channels in the leads that act as spin sinks on time scales much shorter  that the tunneling time, as mentioned already before. 

\section{Tunneling rates and electrical current}
\label{sec3}

\begin{figure}[t]
\begin{center}
\includegraphics[width=0.9\linewidth]{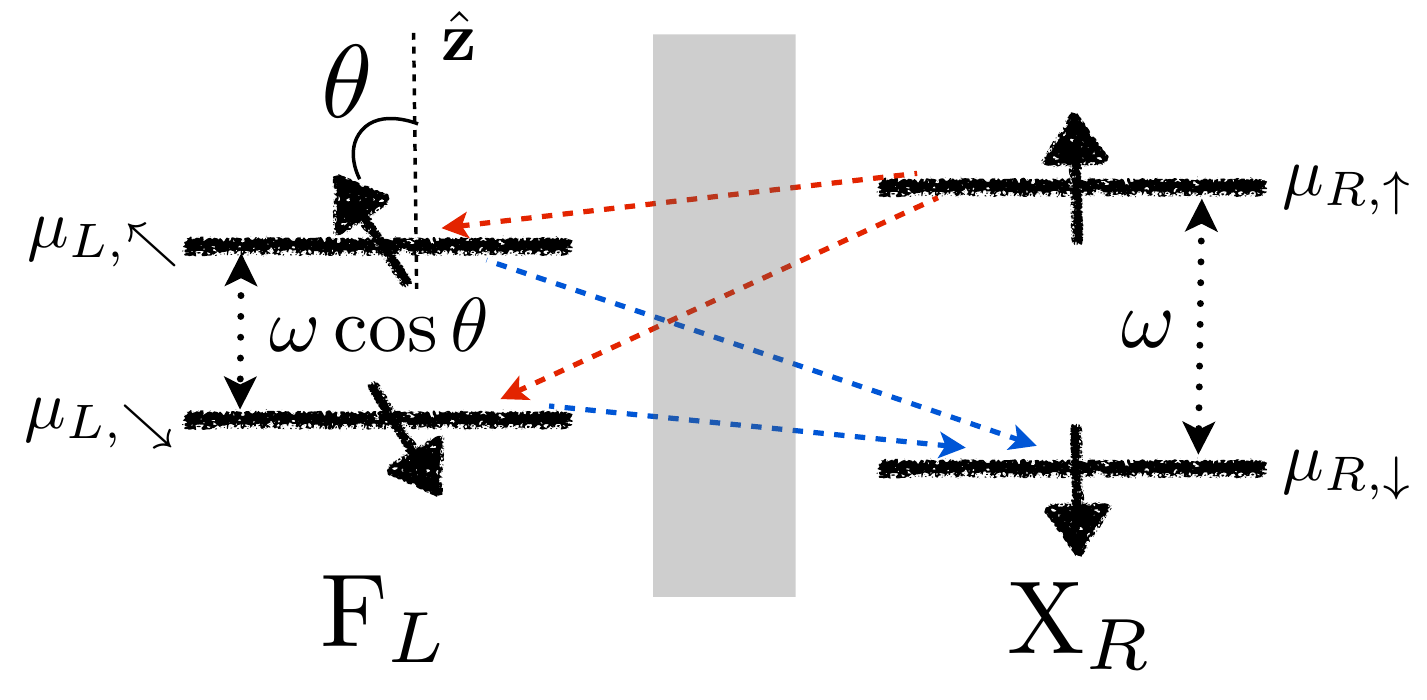}
\caption{The sketch of the level structure in the rotating frame and the electronic transitions for the closed circuit. The left ferromagnet (F$_L$) has its magnetization pointing at an angle $\theta$ with respect to the $\hat{\bm{ z}}$ axis, while the right metal, with  X$_R$=F or N, has the  the spin quantization axis along $\hat{\bm{z}}$. The fictitious magnetic fields leads to a spin splitting by $\omega\cos{\theta}$ and $\omega$ for the left and right metals, respectively. The different spin-dependent chemical potentials on the left ($\mu_{L,\tilde{\sigma}}$) and on the right ($\mu_{R,\sigma}$) metals leads to spin-dependent charge flows between the two parts. The associated voltages are given by $V_{\sigma\sigma'}\equiv\mu_{L,\tilde{\sigma}}-\mu_{R,\sigma'}$, and lead to  different spin transitions mentioned in the text and which are depicted with dashed arrows. } 
\label{Levels_sketch}
\end{center}
\end{figure}

In the tunneling regime, we can compute the electrical current  flowing through the junction by  using the Fermi's golden rule. A sketch of all the tunneling processes and the corresponding fictitious voltages that drive electrons over the tunnel barrier are depicted in Fig.~\ref{Levels_sketch}.  To keep the setup as general as possible, we also  assume that an external  bias $V$ is  applied over the junction (not taken into account in Fig.~\ref{Levels_sketch} for simplicity). The rate from the left to the right lead and from spin $\sigma$ to spin $\sigma'$ (with the corresponding quantization axis in each lead) reads: 
\begin{align}
\Gamma_{\sigma\sigma'}^{LR}(V)&=D^L_{\sigma}D^R_{\sigma'}\int_{-\infty}^{\infty}\int_{-\infty}^{\infty}d\epsilon d\epsilon'|T_{\sigma\sigma'}|^2\nonumber\\
&\times f(\epsilon)[1-f(\epsilon'+eV+eV_{\sigma\sigma'})]P(\epsilon-\epsilon')\nonumber\\
&=D^L_{\sigma}D^R_{\sigma'}\int_{-\infty}^{\infty}d\epsilon|T_{\sigma\sigma'}|^2\frac{\epsilon}{1-e^{-\beta\epsilon}}P(eV+eV_{\sigma\sigma'}-\epsilon)\,,
\label{Gamma}
\end{align}
where $P(E)=\int_{-\infty}^{\infty}dte^{iE t+J(t)}$ is the environment distribution function, the so called $P(E)$ function \cite{DevoretPRL90},  and $V_{\sigma\sigma'}=\omega\,(\sigma\cos{\theta}-\sigma')/2$
represents an effective potential bias acting in the $\sigma-\sigma'$ channel. Here,  
\begin{align}
J(t)=2\int_0^\infty\frac{d\omega}{\omega}\frac{{\rm Re}[Z_t(\omega)]}{R_K}\left[\coth{\left(\frac{\beta\omega}{2}\right)}\left(\cos{\omega t}-1\right)-i\sin{\omega t}\right]\,,
\label{Joft}
\end{align}
with $Z_t(\omega)=1/[i\omega\,C+Z_e^{-1}(\omega)]$ being the total impedance of the circuit, $C$ is the capacitance of the MTJ, $Z_e(\omega)$ is the impedance of the environment, and $R_K=h/e^2$ is the  quantum of resistance.  Note that $V_{\sigma\sigma'}=-V_{\bar{\sigma}\bar{\sigma}'}$, so that $V_{\downarrow\downarrow}=-V_{\uparrow\uparrow}\equiv V_{-}$ and $V_{\uparrow\downarrow}=-V_{\downarrow\uparrow}\equiv V_{+}$, pertaining to $V_{\pm}\geq0$.  We can find the total rate from the left to the right lead by summing over all the spin channels in both leads,  namely  $\Gamma_{LR}=\sum_{\sigma,\sigma'}\Gamma^{LR}_{\sigma\sigma'}$. In order to find the rate from right to left, we can simply use the relation $\Gamma_{RL}(V)=\Gamma_{LR}(-V)$ \cite{IngoldNazarov1990},  which then allows us to write the total current flowing through the junction [and which is given by $I=e(\Gamma_{LR}-\Gamma_{RL})$] as follows:
\begin{align}
I&=\Gamma_{\rm tot}\sum_{s;r=\pm}(1-srP_L)(1+rP_R)\left(1-e^{-\beta(V-rV_s)}\right)T_{s}(\theta)\nonumber\\
&\times\int_{-\infty}^{\infty}d\epsilon \frac{\epsilon}{1-e^{-\beta\epsilon}}
P(V-rV_s-\epsilon)\,,
\label{current_full}
\end{align} 
where $T_{s}(\theta)=\delta_{s,+}\cos^2{(\theta/2)}+\delta_{s,-}\sin^2{(\theta/2)}$,  $\Gamma_{\rm tot}=2\pi|t|^2(D^L_{\uparrow}+D_{\downarrow}^L)(D^R_{\uparrow}+D_{\downarrow}^R)$, and  we defined the following individual polarizations in the left and right leads, respectively, $P_{L,R}=(D^{L,R}_{\uparrow}-D^{L,R}_{\downarrow})/(D^{L,R}_{\uparrow}+D^{L,R}_{\downarrow})$.
The voltage $V$ can be either applied externally, it can be induced by the magnetization dynamics, or can be the sum of the two.

In the open circuit and in the absence of any externally applied voltage, the resulting current vanishes, i.e. $I=0$, so that there is a voltage drop $V$ (see Fig.~\ref{STM_sketch}) induced by the magnetization dynamics. Our aim here is to find precisely this voltage $V$ in the presence of environment. Nevertheless, the current in the closed circuit  is also a measure of the magnetization dynamics. However, in the tunneling regime considered here such a current is very small and hard to detect, as it is proportional to $|t|^2$. The voltage drop, on the other hand, does not depend on the small tunneling parameter $t$, and can reach values comparable to $\omega$,  as shown in the following. Before analyzing in detail the effect of the environment on the charge current, let us first  recover the known results in the absence of it. Such a case corresponds  to  $P(\epsilon)=\delta(\epsilon)$ in the above expression and, assuming also the open circuit setup pertaining to $I=0$, we obtain at zero temperature:
\begin{align}
V=\frac{P_R\,\omega\sin^2{\theta}}{2(1+P_LP_R\cos{\theta})}\,,
\label{bareVoltage}
\end{align}
which indeed corresponds to the result of Tserkovnyak et al. in Ref.~\onlinecite{TserkovnyakPRB08} for the case equivalent ferromagnets ($P_L=P_R=P)$.  Note that if the right ferromagnet is substituted with a normal metal ($P_R=0$) there is no voltage  in the open circuit, which was also shown previously: there is no detection possible in a setup of the form $F|I|N$.  This raises the  question  whether  the presence of environment can lead to  a non-zero voltage when the right ferromagnet becomes a normal metal. To make this more transparent, it is instructive to compute the current in the closed circuit at $P_R=0$ instead of the voltage, which reads:
\begin{align}
I&=\Gamma_{\rm tot}P_L\int_{0}^{\infty}d\epsilon \epsilon\left\{\cos^2{\left(\frac{\theta}{2}\right)}P(V_--\epsilon)-P(V_+-\epsilon)\sin^2{\left(\frac{\theta}{2}\right)}\right\}\,.
\end{align} 
It is easy to see that in the absence of environment this leads to $I=0$, as expected, but in general  this  expression is non-zero. This implies that in the presence of environment the magnetization dynamics detection becomes possible even if the second metal is non-magnetic. The physical picture behind such a non-zero current lies in the fact that the environment, through the inelastic processes, introduces an effective energy-dependent density of states which, in the rotating frame, translates the right normal metal effectively into an ferromagnet.  

\section{Specific environments}
\label{sec4}

To get a better understanding on the effect of an environment, we assume in  the following two particular types of environments, namely a single electromagnetic mode and an  ohmic environment, respectively,  coupled to the MTJ. 

\subsection{Single-mode environment}
\label{sm}

\begin{figure}[t]
\begin{center}
\includegraphics[width=0.99\linewidth]{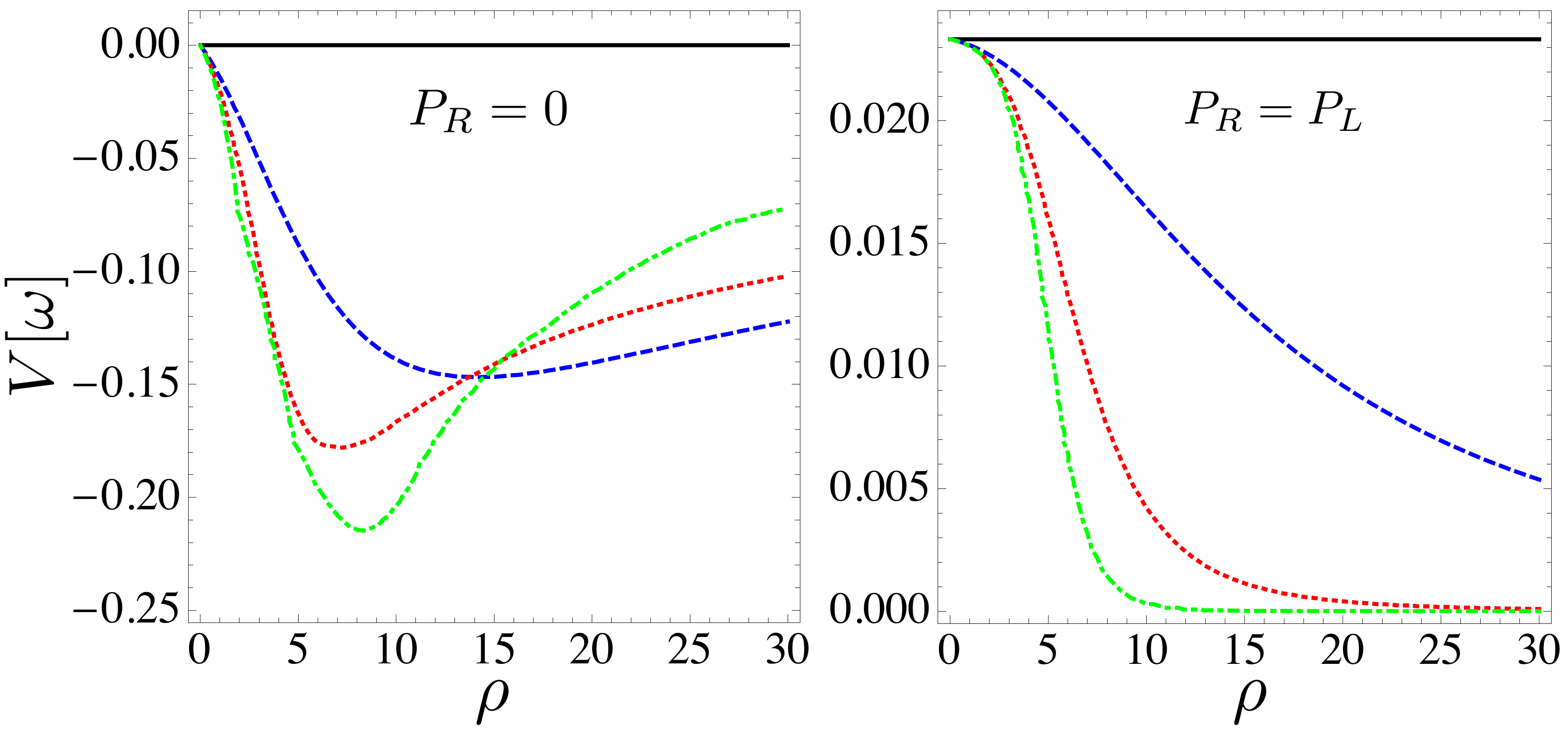}
\caption{The voltage as a function of $\rho$  for a single-mode environment with a normal right lead (left) and equally polarized leads (right). The the black, blue, red, and  green curves, correspond respectively to  $\omega/\omega_0=1$, $3$, $5$ and $10$, with $\theta=\pi/10$. We also consider $T=0$ (zero temperature). }
\label{VoltageRho}
\end{center}
\end{figure}
\begin{figure}[t]
\begin{center}
\includegraphics[width=0.99\linewidth]{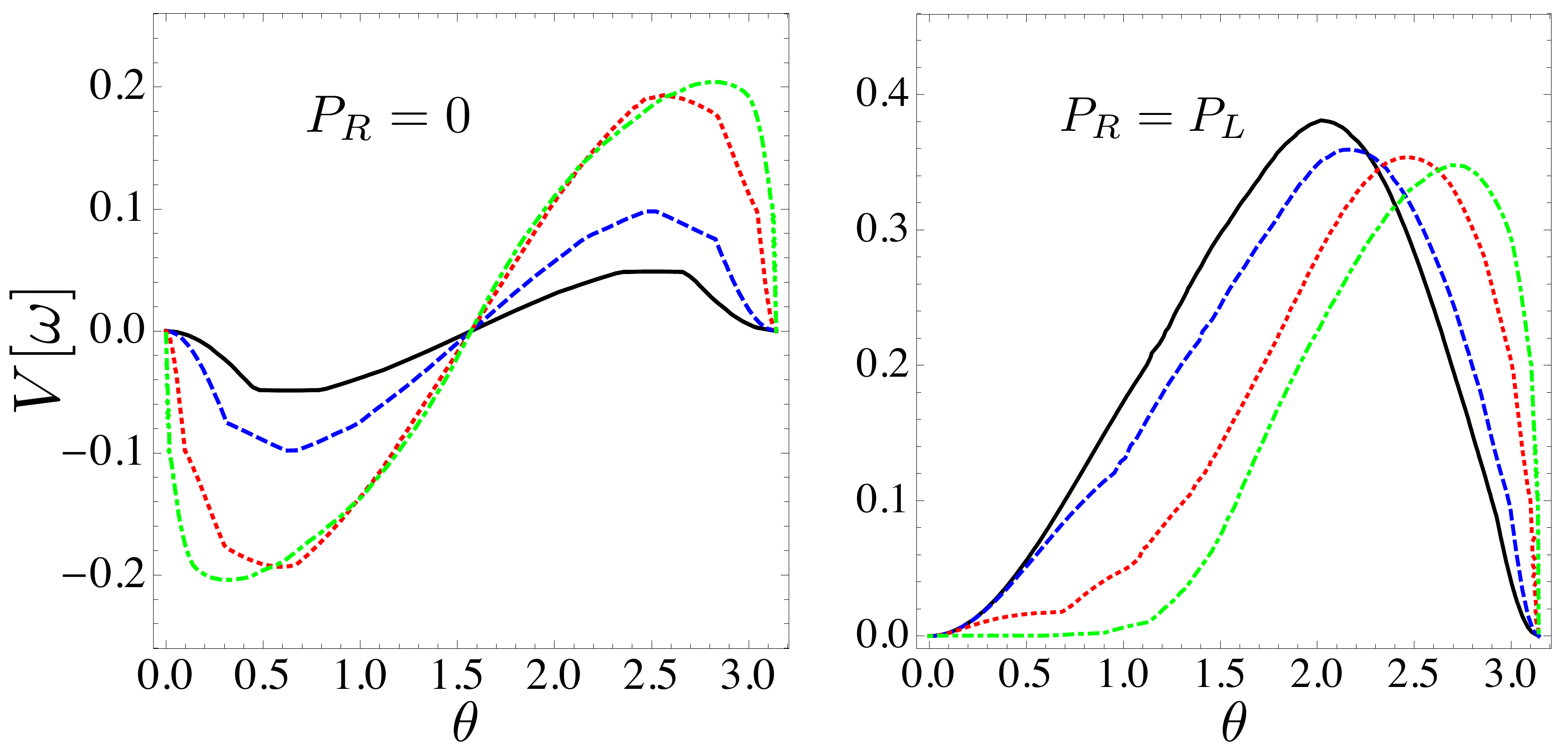}
\caption{The voltage as a function of $\theta$  for a single-mode environment with a normal right lead (left) and equally polarized leads (right). The parameters are as follows:  $\rho=1$, $2$, $5$, and $10$, corresponding to the black, blue, red, and  green curves, respectively. We also consider $T=0$ (zero temperature).}
\label{VoltageTheta}
\end{center}
\end{figure}
\begin{figure}[t]
\begin{center}
\includegraphics[width=0.99\linewidth]{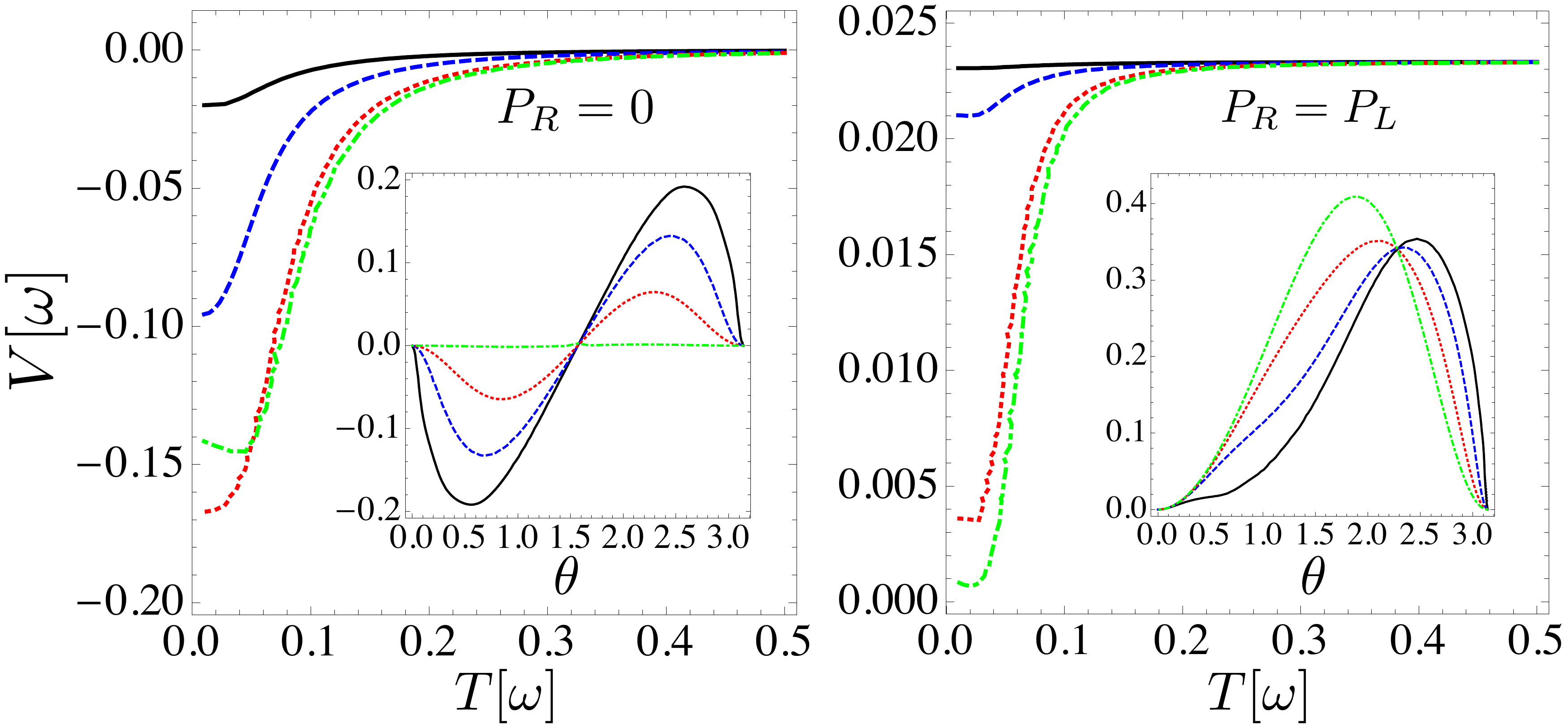}
\caption{The voltage as a function of temperature $T$  for a single-mode environment with a normal right lead (left) and equally polarized leads (right). The parameters are as follows:  $\theta=\pi/10$ and $\rho=1$, $3$, $10$ and $15$, corresponding  to the black, blue, red,  and green curves, respectively. In the insets, we plot the voltage as a function of  $\theta$ for $\rho=5$, and for  $T=0.01$, $0.05$, $0.1$, and $0.5$, corresponding to the black, blue, red, and  green curves, respectively.}
\label{VoltageTemp}
\end{center}
\end{figure}

The single mode environment simply means an  $LC$ circuit, with $C$ being the capacitance of the MTJ (for a sample-STM tip capacitance see, for example, Ref.~\onlinecite{KurokawaJAP98}) and $L$ the inductance of the external circuit. The impedance becomes $Z_e(\omega)=i\omega/[\omega_0^2+(\omega-i\epsilon)^2]$, with $\omega_0=1/\sqrt{LC}$, and $\epsilon$ a small positive number. Such an impedance results in the following distribution function $P(E)$ for the environment at finite temperature:
\begin{equation}
P(E)=e^{-\rho\coth{(\beta\omega_0/2)}}\sum_{n=-\infty}^{n=\infty}I_{n}\left(\frac{\rho}{\sinh{(\beta\omega_0/2)}}\right)e^{n\beta\omega_0/2}\delta(E-n\omega_0)\,,
\end{equation}
where $\rho=(\pi/\omega_0CR_K)=E_c/\omega_0$ is a dimensionless parameter that quantifies the strength of the DCB, i.e. the charging effects versus the environmental excitation energies (here $E_c=e^2/2C$), $\beta=1/k_BT$, and $I_n(x)$ is the modified Bessel function of the first kind. While we cannot extract any analytical expressions for the induced voltage in this case,  in Fig.~\ref{VoltageRho}  we plot the resulting voltage $V$ as a function of $\rho$ by solving the equation $I=0$ for the current in Eq.~\eqref{current_full}, and using  the expression for $P(E)$ at $T=0$. 

The environment strongly affects the induced voltage, and gives rise to a non-monotonic behavior as a function of $\rho$. This  is increased for $P_R=0$ as compared to $P_L=P_R$, reaching values of the order of $\omega$ itself (all energies are expressed in terms of $\omega$). We mention that this turns out to be the opposite for precessions around $\theta=\pi$, as one can see from the left plot in Fig.~\ref{VoltageTheta}. The reason for such a different behavior lies in the fact that the environment affects the spin-dependent density of states $D_{\uparrow}$ and $D_{\downarrow}$  by increasing  (decreasing) their difference  for $\theta=\pi$ ($\theta=0$). More importantly, the environment pertains to large values of $V$ at very small $\theta$, as seen from Fig.~\ref{VoltageTheta}, in total contrast to the environment-free case. Such a surprising feature could be very useful in detection of small-angle magnetization dynamics precession without the need of any reference ferromagnet. 

Next we discuss briefly the influence of thermal fluctuations. At finite temperatures the effect of the environment is diminished, as  depicted in Fig.~\ref{VoltageTemp}. For temperatures $T\ll\omega$ the voltage remains rather unaffected, but as the temperature is increased further,  the voltage is decreased (increased) in the zero polarized lead case $P_R=0$ (equally polarized leads case $P_R=P_L$). Such a behavior is simply due to the fact that thermal excitations wash out the singular spin-polarized density of states induced by the environment.  

%We further proceed by  analyzing another environment,  that will allow us to both demonstrate the generality of the above results and to extract analytical expressions for the induced voltage.   

\subsection{Ohmic environment}
\label{ohmic}

As the next example, let us analyze the MTJ in the presence of an Ohmic environment.  For such a case, the impedance reads \cite{DevoretPRL90,IngoldNazarov1990}:
\begin{equation}
\frac{{\rm Re}[Z_t(\omega)]}{R_K}=\frac{1}{g}\frac{1}{1+(\omega/\omega_R)^2}\,,
\end{equation}
where $g=R_K/R$, and  $\omega_R=1/RC\equiv (g/\pi)E_c$. At finite temperatures, we can write the the full correlation function in Eq.~\eqref{Joft} as $J(t)=J_{\rm q}(t)+J_{\rm th}(t)$, with $J_{\rm q}(t)$ and $J_{\rm th}(t)$ being the quantum (or the $T=0$) and the thermal contributions, respectively.  Here we are interested in the behavior of the junction at energies well below the charging energy $E_c$, where possible zero-bias anomalies are manifesting. This allows us to consider only the long time limit $t\rightarrow\infty$ characteristic of the correlation function $J(t)$. More specifically, we assume $\omega_R t\gg1$, which leads to the following expressions:
\begin{align}
J_{\rm q}(t)&\approx-\frac{2}{g}\left[\log{(\omega_R t)}+i\frac{\pi}{2}+\gamma\right]\,,\\
J_{\rm th}(t)&\approx\frac{2}{g}\log{\left(\frac{\pi|t|\, T}{\sinh{(\pi|t|\,T)}}\right)}\,,
\label{Johmic}
\end{align} 
with  $\gamma=0.577$ being the Euler constant. We note that for $T\,t\gg1$, the thermal contribution becomes $J_{\rm th}(t)\approx-\pi T\,t/g$, \cite{DevoretPRL90} while at short times $T\,t\ll1$ (but still  with $\omega_Rt\gg1$), the correlation function depends quadratically on time, i.e. $J_{\rm th}(t)\approx-\pi^2 T^2\,t^2/6g$.  The full correlation function $J(t)$ pertains to an exact analytical expression for the $P(E)$ function defined after Eq.~\eqref{Gamma}, and  which characterizes the probability to excite the environment at energy $E$. However, we will not  show the full  expression since it is too lengthy to be displayed. Instead, we depict the $T=0$ expression for $P(E)$, which reads \cite{DevoretPRL90}:
\begin{equation}
P(E)=\frac{e^{-2\gamma/g}}{\Gamma(2/g)}\frac{1}{E}\left(\frac{\pi}{g}\frac{E}{E_c}\right)^{2/g}\,,
\label{PEohmic0}
\end{equation}
\begin{figure}[t]
\begin{center}
\includegraphics[width=0.9\linewidth]{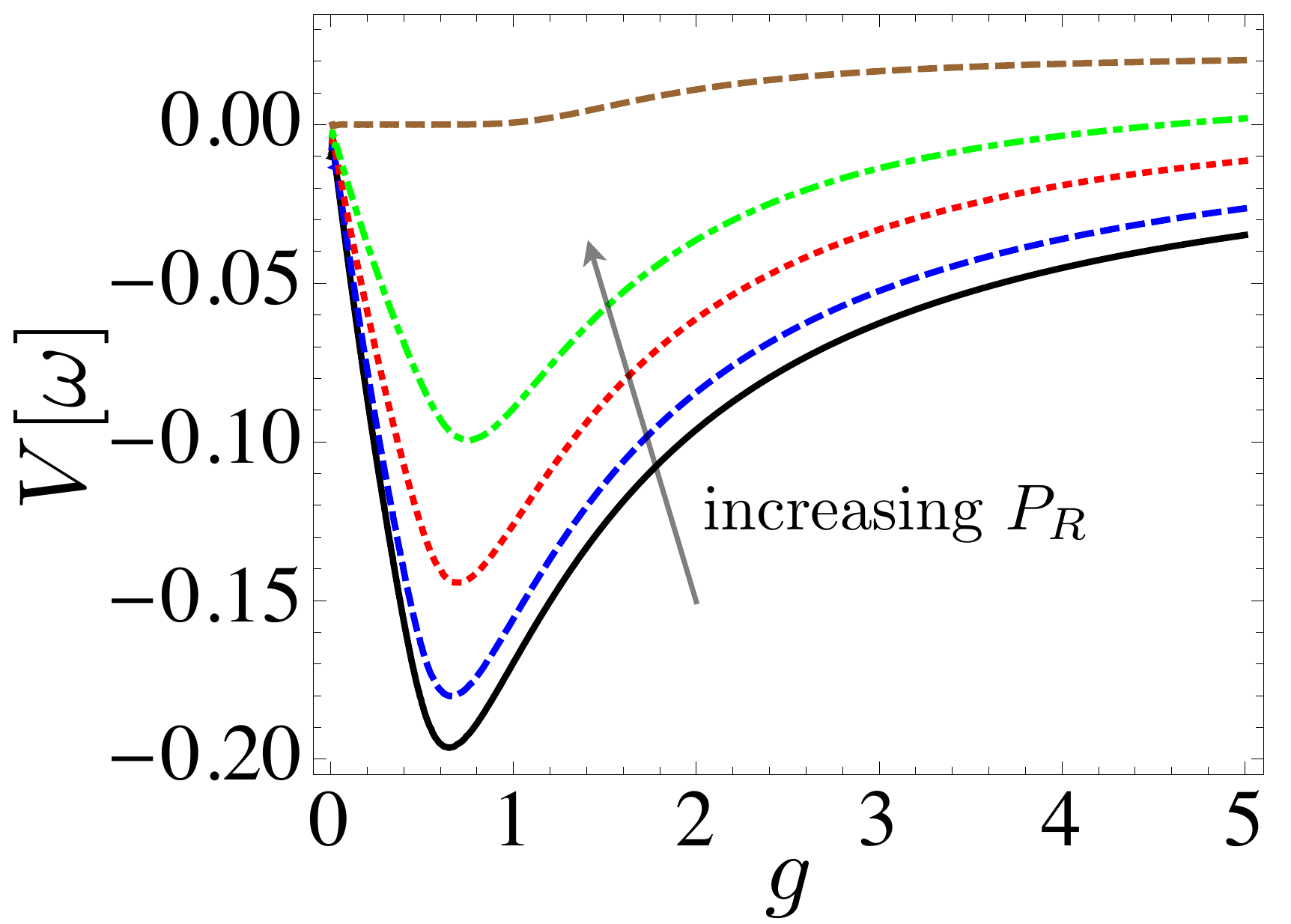}
\caption{The voltage as a function of the coupling parameter $g$ for  an ohmic environment at $T=0$. The plot parameters are as follows:  $\theta=\pi/10$ and $P_R=0$, $0.1$, $0.3$, $0.5$, and  $0.75$, corresponding respectively to the black, blue, red, green,  and brown curves.}
\label{VoltageOhmicG}
\end{center}
\end{figure}
$\Gamma(x)$ being  the Gamma function. 

\begin{figure}[t]
\begin{center}
\includegraphics[width=0.95\linewidth]{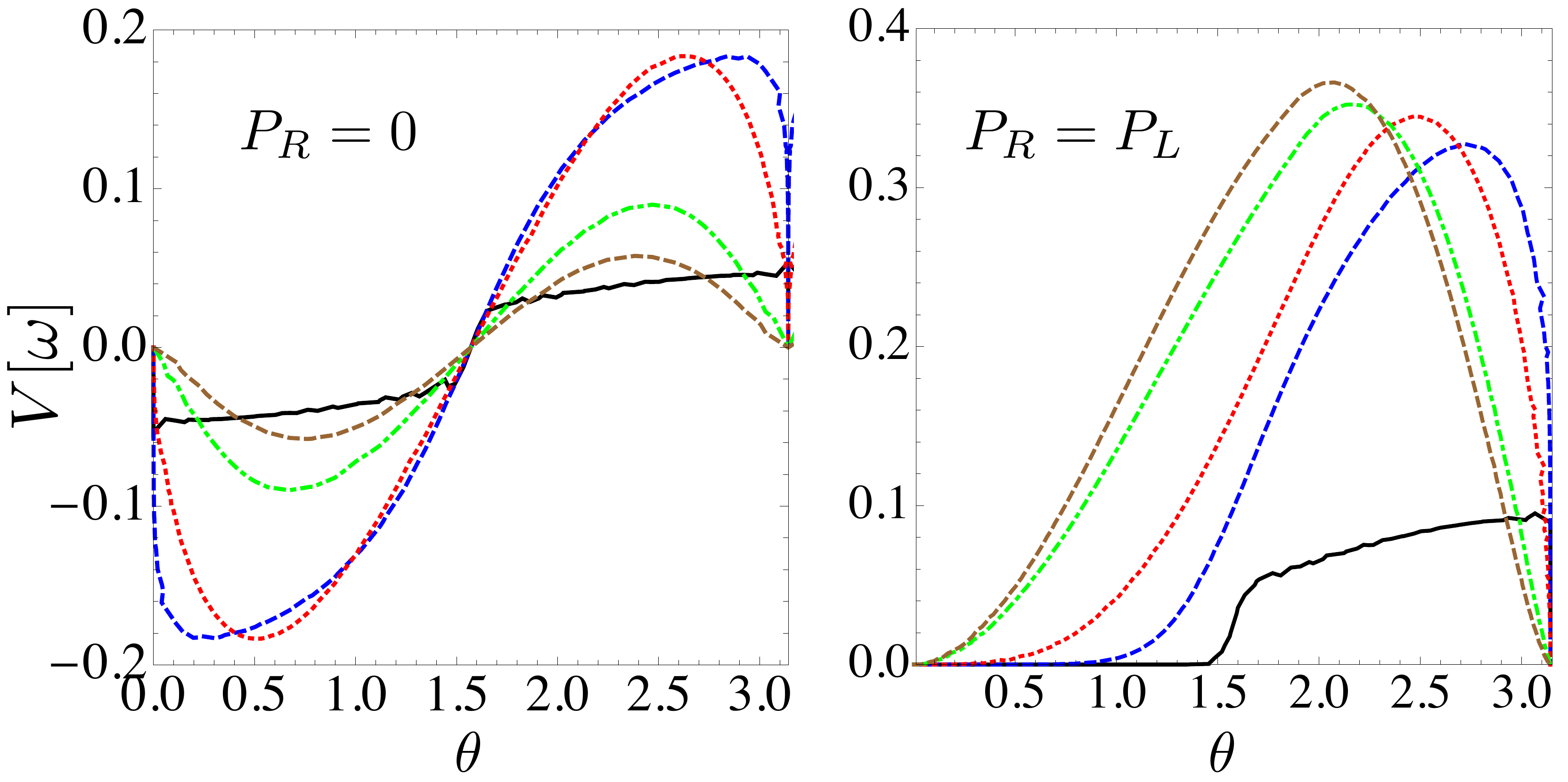}
\caption{The voltage as a function of  $\theta$  for an ohmic environment at $T=0$. The left plot shows the $P_R=0$ case, while the right plot shows  the voltage for the case of  equivalent ferromagnets $P_R=P_L$. The  black, blue, red,  green, and brown curves correspond, respectively to $g=0.1$, $0.5$, $1$, $3$, and $5$.}
\label{VoltageOhmicTheta}
\end{center}
\end{figure}

Similarly to the the single-mode case, to  find the voltage induced by the magnetization dynamics we use Eq.~\eqref{current_full} and solve the equation $I=0$  valid for the open circuit.  In Fig.~\ref{VoltageOhmicG} we plot the resulting voltage at $T=0$ [using Eq.~\eqref{PEohmic0}] as a function of  the ohmic parameter $g$ and for several polarizations of the right lead $P_R$, assuming a precession cone angle $\theta=\pi/10$.  We see that the voltage is non-monotonic as a function of the ohmic parameter $g$, reaching a maximum around $g\sim1$ (or $R\sim R_K$). Moreover, the voltage decreases with increasing the polarization $P_R$ of the right lead, meaning that the detection scheme in the presence of the environment  is more efficient with a normal metal as a detector instead of a ferromagnet as in the original setup \cite{TserkovnyakPRB08}.  

\begin{figure}[t]
\begin{center}
\includegraphics[width=0.9\linewidth]{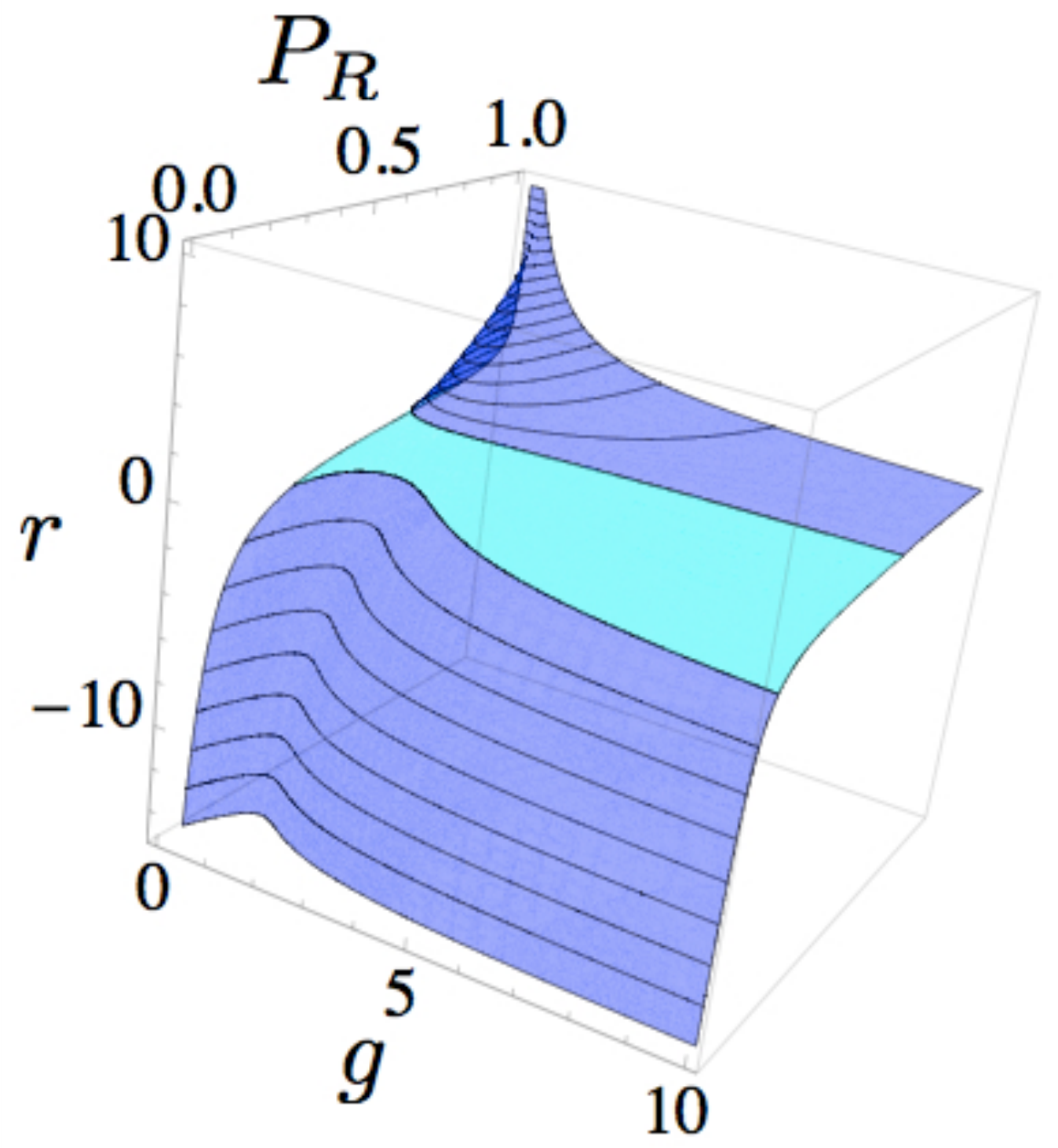}
\caption{The amplification function $r$ as a function of the ohmic parameter $g$ and the right lead polarization $P_R$, for a precession cone angle $\theta=\pi/10$ and at $T=0$. In blue it is depicted  the region for which $|r|>1$, while in light blue we show the region for which $|r|<1$.  Note that the amplification factor $r$ can be  both positive and negative over a wide range of parameters.}
\label{VoltageAmpl}
\end{center}
\end{figure}

In Fig.~\ref{VoltageOhmicTheta} we plot the $T=0$ voltage as a function of the cone angle $\theta$ for several values of the Ohmic coupling parameter $g$,  $P_R=0$ (left plot) and $P_R=P_L$ (right plot). Similarly to the previous case, the voltage reaches values of the order $\omega$ close to $\theta=0$, again in stark contrast to the environment-free case. Note that the absence of the environment corresponds to $g\rightarrow\infty$, in which case the voltage is zero at  $P_R=0$. While the features are very similar to the single-mode environment, this case allows us to extract analytical expressions for the induced voltage, in the limit of small precession angles $\theta$. For the case of a $F|I|N$ structure, the polarization of the right lead is $P_R\equiv0$, and we find the following asymptotic expressions ($V$ is expressed in units of $\omega)$:
\begin{align}
V&=\left\{
\begin{array}{cc}
\displaystyle{-\frac{4.25P_L\sin^2{(\theta/2})}{g}}\,, & {\rm for}\,\, g\rightarrow\infty\\\\
\displaystyle{-\frac{gP_L}{2}}\,, & {\rm for}\,\, g\rightarrow0
\end{array}
\right.\,,
\label{theta0result}
\end{align}
meaning that the voltage approaches zero as $1/g$ as the environment strength is diminished. For a strong environment coupling ($g\rightarrow0$), the voltage increases linearly with $g$, and becomes independent of the angle $\theta$.  Such a behavior, however,  is only  valid for finite cone angles $\theta$, while for $\theta\rightarrow0$ the voltage vanishes  as expected (and as shown in Fig.~\ref{VoltageOhmicTheta} where the exact expression is plotted). Note that the voltage around $\theta=\pi$ is identical to the $\pi=0$, but with opposite sign. In the  limit of identical ferromagnets on both sides ($P_L=P_R\equiv P$), we obtain  
\begin{align}
V&=\left\{
\begin{array}{cc}
\displaystyle{\frac{2P\sin^2{(\theta/2)}}{1+P^2}\,\,\,\, \left(-\frac{2P\sin^2{(\theta/2)}}{1-P^2}\right)}\,, & {\rm for}\,\, g\rightarrow\infty\\\\
\displaystyle{\frac{gP}{1-P^2}\sin^{4/g}{(\theta/2)}\,\,\,\, \left(-\frac{gP}{1+P^2}\right)}\,, & {\rm for}\,\, g\rightarrow0
\end{array}
\right.\,.
\label{theta0result}
\end{align}
corresponding to angles $\theta\approx0$, while the expression in the brackets represents the expansions around $\theta\approx\pi$. We see that for $\theta\approx0$ ($\theta\approx\pi$) the voltage is strongly reduced (enhanced) compared to its bare values in the absence of the environment ($g\rightarrow\infty$). As before,  such a behavior is again due to the effective environmental  density of states which tends to reduce (enhance) the difference between $D_\uparrow$ and $D_{\downarrow}$ around $\theta=0$ ($\theta=\pi$). 

While the density of states modification by the environment offers an explanation for the non-zero voltage in the $F|I|N$ case, it does not explain why this voltage reaches a maximum comparable to $\omega$ around $\theta=0$ (or $\theta=\pi$). For a qualitative explanation, we make use of the level structure and electronic transitions depicted in  Fig.~\ref{Levels_sketch}. We see that the transitions are between energy levels separated by $\delta E=\omega(1\pm\cos{\theta})$, which means the environment is interrogated  (through the $P(E)$ function) at these two energy scales. When $\theta\rightarrow0$ (or $\theta\rightarrow\pi$), the environment is probed at both $E\approx\omega$ and $E\approx0$. Close to $E\approx0$, the zero bias anomaly sets in,  which reflects a very resistive environment and thus some of the channels for the back-flow charge currents are highly inhibited, leading to an amplification of the voltage.

Next  we characterize the voltage gain in the presence of the environment compared to the environment-free case. For that, we define the amplification factor $r$ as:
\begin{equation}
r=\frac{V(g)}{V(\infty)}\,,
\end{equation}
where $V(\infty)$ is the voltage in the absence of environment in Eq.~\eqref{bareVoltage}. In Fig.~\ref{VoltageAmpl} we plot $r$ as a function of $g$ and $P_R$ for a precession cone angle $\theta=\pi/10$. We see that  $|r|>1$, or even $|r|\gg1$, for a large range of parameters, and thus the detection scheme improves substantially in the presence of environment. Moreover, the amplification factor $r$ can run over both positive and negative values, meaning the voltage  can reverse sign, a feature associated again with the effective ferromagnetic density of states induced by environment.  

\begin{figure}[t]
\begin{center}
\includegraphics[width=0.95\linewidth]{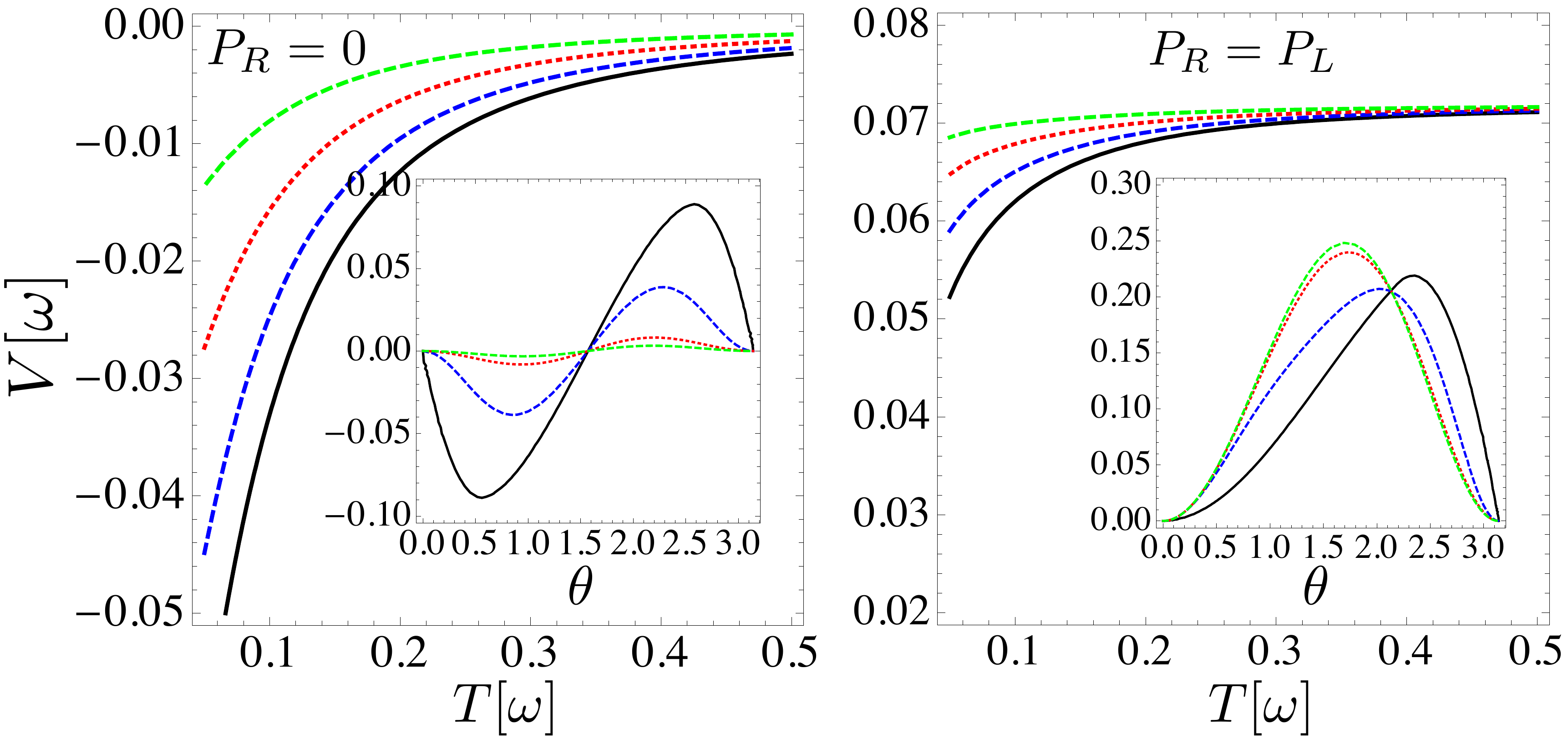}
\caption{The voltage as a function of temperature $T$ and angle $\theta$ for a normal metal right lead (left plot) and a ferromagnetic right lead (right plot) for an ohmic environment. The black, blue, red, and green curves correspond  to $\theta$ for $g=2.1$, $3$, $5$, $10$, respectively. In the insets we plot the angular dependence of the induced voltage for different temperatures $T$: the black, blue,  red, and green curves correspond to $T=0$, $0.1$, $0.3$, and $0.5$, respectively.}
\label{VoltageOhmicTemp}
\end{center}
\end{figure}

As in the previous case, the thermal fluctuations are expected to smear out the sharp features induced by the the coupling to the environment. However, for small enough temperatures so that $T\ll\omega$, the main features remain visible, as shown in Fig.~\ref{VoltageOhmicTemp}. There, we plot the induced voltage as a function of temperature for different ohmic parameters $g$. More specifically, in the main left plot we show the result for $P_R=0$, while in the main right plot  we depict the result for $P_R=P_L$.  We mention that  in the first case the voltage vanishes as the temperature is increased, while in the second case the voltage reaches the environment-free value for large temperatures, as the singular features are washed out by the thermal fluctuations. In the insets we show  the angular dependence of the voltage for different temperatures. The sharp features around $\theta=0$, $\pi$ are pushed to larger cone angles $\theta$ compared to the $T=0$ case, thus reducing the efficiency of the detection scheme (since usually the cone angles are rather small, $\theta<5$ rad).  

Finally, let us give some realistic estimates for the induced voltage based on numbers utilized for observing the DCB in normal (ultra-small) tunnel junctions \cite{PierreNatPhys11,PierreNatComm12} and  Josephson (superconducting) junctions \cite{PortierPRL11}. Since we are considering ferromagnetic metals, we believe that such estimates are also appropriate for ultra-small MTJs, such as those investigated here.  For example, in Ref.~\onlinecite{PierreNatPhys11} it was reported that $Z_{e}(\omega)\approx R\sim R_K$ up to frequencies of the order $\omega\sim1$ GHz,  $C\approx 2$ fF, which translates to $E_C\approx 10^{-4}$ eV, and $T=25$ mK, so that $E_C,\omega \gg k_BT$. These numbers  give $g\sim1$ and lead for a large value of the induced voltage, $eV\approx0.2\,\hbar\omega$. More conservative numbers, such as those in Ref.~\onlinecite{PierreNatPhys11} (tunnel junction) and Ref.~\onlinecite{PortierPRL11} (Josephson junction) lead to values of $g$ between $g=0.1$ and $g=0.4$, which in turn pertains  to a voltage $eV\sim0.01-0.15\,\hbar\omega$ at small precession angle $\theta=\pi/10$ and for a normal metal as a right lead (the result is $eV=0$ in the absence of environment in such a case).       

\section{Conclusions}
\label{sec5}

In this paper we studied the magnetization dynamics induced voltage in an ultra-small magnetic tunnel junction. We show that in the presence of an electromagnetic environment the electronic transport shows  a singular behavior which in turn affects dramatically the induced voltage characteristic. More specifically, we find that the voltage pertaining to the magnetization dynamics can be comparable to the precession frequency $\omega$ even for small precessions angles $\theta\sim0$, being in stark contrast with behavior of a large MTJ where such a voltage scales as $V\propto\sin^2{\theta}$. Moreover, the resulting voltage is finite even when the reference ferromagnet becomes a simple metal, lifting the requirement of using  ferromagnets for detection.   

The above findings could be extremely useful in view of magnetization dynamics applications. It could provide  the means for detecting the  magnetization dynamics, for example, on the surface of a 2D ferromagnet by usual STM technique without the need for a spin polarized reference (i.e. spin polarized tip). Turning the tables, the small-angle magnetic dynamics can serve as a nonintrusive spectroscopic probe of the  environment itself. By  addressing the voltage as a function of $\omega$ one can  characterize the nonlinear properties of the environment, possibly even prepared in some given state, or characterize the  tunneling into the many-body system itself. 

We thank Yaroslav Tserkovnyak and Andre Thiaville for helpful discussions and suggestions. This work is supported by a public grant from the ``Laboratoire d'Excellence Physics Atom Light Mater" (LabEx PALM, reference: ANR-10-LABX-0039).

\end{document}